# White Paper

# EFSPI CMCSNE SIG position on the ‘Expected $f_2$’

**Participants of Working Group**

**Contributors**

- Thomas Hoffelder, Senior Principal Statistician, Boehringer Ingelheim Pharma GmbH & Co. KG, Global Biostatistics & Data Sciences, Binger Str. 173, 55216 Ingelheim am Rhein, Germany
- Marijn Clement, Senior Statistician, N.V. Organon, Center for Mathematical Sciences, Kloosterstraat 6, 5349 AB Oss, Netherlands
- Stephen Karanja, Senior Statistician, Johnson & Johnson, Discovery and Manufacturing Statistics, Turnhoutseweg 30, 2340 Beerse, Belgium
- Marta Regis, Assistant Professor in Applied Statistics, Technische Universiteit Eindhoven, Het Eeuwsel 53, 5612 AZ Eindhoven, Netherlands
- Marc Lindenberg, Senior Principal Scientist, F. Hoffmann-La Roche AG, Galenical & Analytical Development, Grenzacher Strasse 124, 4070 Basel, Switzerland
- Filip Vestin, Senior Statistician, Novo Nordisk A/S, Specialists & Adv. Data Analytics, Girostrøget 1, 2630 Taastrup, Denmark
- Eleonore Pablik, Principal TAM Stat, Boehringer Ingelheim Pharma GmbH & Co. KG, Global Biostatistics & Data Sciences, Binger Str. 173, 55216 Ingelheim am Rhein, Germany
- Marion Berger, Statistical Pharmaceutical Development Team Leader, Sanofi S.A., Computational and AI Strategy-Data Sciences, 371 Rue du Professeur Joseph Blayac, 34184 Montpellier Cedex04, France
- Valgeir Einarsson, Statistician, Novo Nordisk A/S, Specialists & Adv. Data Analytics, Girostrøget 1, 2630 Taastrup, Denmark
- Kevin Lief, Chair of CMC Statistical Network Europe SIG, EFSPI & Statistics Director, GlaxoSmithKline, CMC Statistical Sciences, Drug Development and Supply, David Jack Centre for R&D, Ware, SG12 0DP, Hertfordshire, UK
- Jens Lamerz, Senior Principal Statistical Scientist, quanthansa.ai, Abtsgasse 19, 77654 Offenburg, Germany

**Observer**

- Helen Thomas, Quality Assessor, Swissmedic, Zulassung und Vigilance Arzneimittel, Hallerstrasse 7, 3012 Bern, Switzerland

## ABSTRACT

The method called 'expected $f_2$' ($\hat{f}_{2,\text{exp}}$), as proposed by Noce et al. (2020) and Xu et al. (2021), has been adopted in two health authority guidelines for dissolution profile comparison when variability precludes the use of the conventional similarity factor $\widehat{f_2}$. This position paper, developed by a working group of the European Federation of Statisticians in the Pharmaceutical Industry CMC Statistical Network Europe Special Interest Group (EFSPI CMCSNE SIG) comprising 12 CMC statisticians from seven pharmaceutical companies[1] and one statistician from academia, with one observer from a regulatory authority, presents a critical evaluation of this method[2]. Fundamental concerns are identified. First, the formula for $\hat{f}_{2,\text{exp}}$ has no traceable origin in the references cited by its proponents. Noce et al. (2020) and Xu et al. (2021) attribute $\hat{f}_{2,\text{exp}}$ to Shah et al. (1998) and Ma et al. (1999, 2000), but a review of these sources shows that neither mentions nor suggests the $\hat{f}_{2,\text{exp}}$ formula. Second, no mathematical justification has been provided for the formula. Where Shah et al. (1998) subtract a variance term to reduce the upward bias of $\widehat{f_2}$, the $\hat{f}_{2,\text{exp}}$ formula adds this term, thereby increasing rather than correcting the bias. This has also been noted by FDA statisticians Liu et al. (2024). Third, the method exhibits poor statistical properties: for highly variable profiles, the variance term dominates the statistic, resulting in low power even as the true difference between profiles approaches zero. The method can reject equivalence when profiles are identical. Fourth, the formula as published by Noce et al. (2020) contains a notation ambiguity that renders the intended grouping of terms unclear. This ambiguity has propagated into regulatory guidance.

A survey of working group members, designed to elicit arguments both for and against the method, found no scientifically meaningful advantage.

The EFSPI CMCSNE SIG concludes that $\hat{f}_{2,\text{exp}}$ should not be recommended for dissolution profile comparison.

---

[1] *Contributors who are employees of pharmaceutical companies may hold shares and/or stock options in their respective company.*

[2] *Their perspectives helped strengthen the clarity and accuracy of the analysis. All contributions by public servants were provided in a personal capacity, and the views expressed in this paper, as well as any remaining errors or omissions, are solely those of the authors and do not represent the views or positions of any government agency or institution.*

# 1 Background and Procedure of the Initiative

In January 2026, the EFSPI CMCSNE SIG issued a call for contributions to all members of the network, from over twenty pharmaceutical companies, inviting them to participate in a systematic evaluation of the $\hat{f}_{2,\text{exp}}$ method for dissolution profile comparison. The initiative was motivated by the growing adoption of $\hat{f}_{2,\text{exp}}$ following its recommendation in EMA's Q&A 3.13 (2025) and aimed to develop a balanced position on the method's advantages, disadvantages and practical experiences. A working group was established at a kick-off meeting on 5 February 2026, comprising 14 volunteers: 12 CMC statisticians from seven pharmaceutical companies, one from academia, and one observer from the Swiss regulatory authority Swissmedic. The working group was chaired on behalf of CMCSNE SIG. Participants were explicitly invited to present arguments both in favour of and against the method, and multiple opportunities were provided to express supportive views through presentations, group discussion and an anonymous survey on perceived advantages and disadvantages. The following sections outline the theoretical and practical considerations underlying $\hat{f}_{2,\text{exp}}$, the survey and discussions through which the working group arrived at its consensus and the recommendations derived from it.

# 2 Motivation for Dissolution Testing

Dissolution tests are among the most widely used tests in the characterisation and quality control of solid oral dosage forms. Classified as performance tests in pharmacopoeia, they measure the release of the active pharmaceutical ingredient, reported as percentage of the nominal dose at predefined time points, yielding dissolution profiles that potentially link in vitro behaviour to in vivo performance. Comparison of dissolution profiles is a statistical topic from the CMC area and serves as a surrogate for in vivo studies to demonstrate equivalence of quality in post-approval changes (such as scale-ups or manufacturing site and equipment changes) and biowaiver requests. In these studies, the manufacturer must demonstrate that the dissolution profile of the product population before the change (REF) is equivalent to that after the change (TEST).

# 3 Notation and definitions

Let $p$ be the number of dissolution time points and $n$ be the sample size of the REF and TEST samples. Note that the method discussed in this paper, $\hat{f}_{2,\exp}$, is only defined for equal sample sizes of both populations under comparison. Let $\boldsymbol{\mu_R} = (\mu_{R1}, \mu_{R2}, \dots, \mu_{Rp})$ and $\boldsymbol{\mu_T} = (\mu_{T1}, \mu_{T2}, \dots, \mu_{Tp})$ represent the expected values of the dissolution profile of REF and TEST population, respectively. $\boldsymbol{\mu_R}$ and $\boldsymbol{\mu_T}$ are p-dimensional vectors. $\boldsymbol{\bar{X}_R} = (\bar{X}_{R1}, \dots, \bar{X}_{Rp})$ is the mean vector of the measured REF profiles and $\boldsymbol{\bar{X}_T} = (\bar{X}_{T1}, \dots, \bar{X}_{Tp})$ is the mean vector of the measured TEST profiles. Let $\boldsymbol{\sigma_R^2} = (\sigma_{R1}^2, \sigma_{R2}^2, \dots, \sigma_{Rp}^2)$ be the vector of variances of the REF population, and $\boldsymbol{\sigma_T^2} = (\sigma_{T1}^2, \sigma_{T2}^2, \dots, \sigma_{Tp}^2)$ be the vector of variances of the the TEST population. Let $\boldsymbol{S_R^2} = (S_{R1}^2, S_{R2}^2, \dots, S_{Rp}^2)$ be the corresponding estimated variances of the REF sample, and $\boldsymbol{S_T^2} = (S_{T1}^2, S_{T2}^2, \dots, S_{Tp}^2)$ be the estimated variances of the TEST sample.

The generalized mean of $p$ numbers $d_1, d_2, \dots, d_p$ is defined as $\left(\frac{1}{p}\left(d_1^a + d_2^a + \dots + d_p^a\right)\right)^{1/a}$.

For $a = 2$, the quantity $\sqrt{\frac{1}{p}(d_1^2 + d_2^2 + \dots + d_p^2)}$ is the “quadratic mean”.

# 4 The similarity factor $\hat{f}_2$

The method recommended in all guidelines for dissolution profile equivalence analyses is the “similarity factor” $\hat{f}_2$ based on the Euclidean distance (ED) $ED = \sqrt{\sum_{i=1}^{p}(\mu_{Ri} - \mu_{Ti})^2}$ between the expected values of REF and TEST. The Euclidean distance is a non-standardized distance measure as the (squared) differences at all time points are summed up without standardization.

The number of time points that are relevant for dissolution profile comparisons depends on the product tested. In general, ED increases with an increasing number of time points. Assuming constant shift between the means, to quantify the dissimilarity independent of the number of time points, one needs to relate ED to the number of time points $p$: let $QMD = \sqrt{(1/p)\sum_{i=1}^{p}(\mu_{Ri} - \mu_{Ti})^2}$ be the quadratic mean (see definition above) of the differences between both true profile means. QMD relates the ED to the number of dissolution time points ($QMD^2 = 1/p\ ED^2$).

The similarity factor $\widehat{f_2}$ is a point estimate of the parameter $f_2$, which is sometimes called "population $f_2$" or "true $f_2$". The parameter $f_2$ is based on a series of monotone transformations of QMD and ED:

$$50 \cdot log_{10}\left(\frac{100}{\sqrt{1+\frac{1}{p}ED^2}}\right) = f_2 = 50 \cdot log_{10}\left(\frac{100}{\sqrt{1+QMD^2}}\right).$$

The point estimates $\widehat{f_2}$, $\widehat{ED}$ and $\widehat{QMD}$ are obtained by replacing the true means $\boldsymbol{\mu_R}$ and $\boldsymbol{\mu_T}$ by their point estimates $\boldsymbol{\bar{X}_R}$ and $\boldsymbol{\bar{X}_T}$. The acceptance criterion "$\widehat{f_2} > 50$" (EMA, 2010; FDA, 1997, among others) is identical to the criterion "$\widehat{QMD} < 9.95 \approx 10$" (or to "$\widehat{ED}^2 < 99p$"). This means that the profile comparison is successful if the average (i.e. the quadratic mean) difference between the profile means is less than 10 (% of label claim). Please note that in some guidelines the acceptance criterion $\widehat{f_2} \geq 50$ can also be found.

The corresponding statistical null hypotheses of non-similarity are:

$H_0$: $f_2 \leq 50$ $H_0$: $QMD \geq 9.95 \approx 10$ $H_0$: $ED^2 \geq 99p$.

**Equation [1]: Corresponding statistical null hypotheses of non-similarity.**

All three variants of $H_0$ are identical. The corresponding alternative hypotheses (of equivalence) are:

$H_a$: $f_2 > 50$ $H_a$: $QMD < 9.95 \approx 10$ $H_a$: $ED^2 < 99p$.

Decisions based solely on the point estimates $\widehat{f_2}$, $\widehat{ED}$ and $\widehat{QMD}$ do not allow Type I Error (T1E) control. Therefore, guidelines allow the use of $\widehat{f_2}$ only in case of low variability (the definition of "low variability" might slightly differ between guidelines). Regional guidelines generally do not provide a concrete recommendation for the statistical method in case of high variability.

This paper shows that "expected $f_2$" is a scientifically questionable and unjustified approach which can lead to underpowered dissolution profile studies. The use of this method can result in high Type II error decisions against profile equivalence and consequently (because of the failed dissolution profile equivalence assessment) in the conduction of unnecessary in vivo bioequivalence trials. We see here a conflict to the ICH E6 and ICH E8 requirements, e.g. “care should be taken to ensure all study procedures and assessments are necessary from a scientific viewpoint and do not place undue burden on study participants”.

Several alternative statistical methods for dissolution profile comparison under high variability have been proposed in the literature, including various bootstrap approaches for $\widehat{f}_2$, asymptotic tests based on the Euclidean distance, and methods based on other distance measures. A comprehensive discussion of these alternatives is beyond the scope of this paper. Here, we focus on the 'expected $f_2$', which has gained regulatory traction through its inclusion in the EMA Q&A 3.13 (2025) and the South Korean MFDS guideline (2023).

## 5 Origin of the 'expected $f_2$' formula and relationship to suggestions of Shah et al. (1998)

A method to check whether the quadratic mean of the profile mean differences is below 10% (of label claim) is via the acceptance criterion:

$$\widehat{QMD} < 9.95 \approx 10 \Leftrightarrow \qquad \widehat{f}_2 = 50 \cdot log_{10}\left(\frac{100}{\sqrt{1+\widehat{QMD}^2}}\right) > 50.$$

Another expression for $\widehat{f}_2$ is:

$$\widehat{f}_2 = 100 - 25 \cdot log_{10}\left(1 + \widehat{QMD}^2\right) = 100 - 25 \cdot log_{10}\left(1 + (1/p)\sum_{i=1}^{p}(\bar{X}_{Ti} - \bar{X}_{Ri})^2\right).$$

For dissolution profile comparison, the first suggested use of the point estimate $\widehat{f}_2$ was in Moore and Flanner (1996), before it percolated into regulatory guidance. The corresponding equivalence hypotheses were not defined neither by Moore and Flanner (1996) nor by the guidances. Shah et al. (1998) searched for a parameter which could be the parameter of interest behind the point estimate $\widehat{f}_2$. They replaced the mean values in $\widehat{f}_2$ by the corresponding expected values and called the resulting parameter

**“Population $f_2$”:** $f_2 = 100 - 25 \cdot log_{10}\left(1 + (1/p)\sum_{i=1}^{p}(\mu_{Ri} - \mu_{Ti})^2\right)$

With this parameter, the statistical hypotheses presented in Equation [1] could be defined. Shah et al. (1998) analysed that $\widehat{f}_2$ can be seen as a biased point estimate for the “Population $f_2$”.

Under the assumption of equal sample sizes for REF and TEST (often but not always fulfilled in dissolution profile studies) the expected value of $\widehat{f_2}$ was **approximated** by Shah et al. (1998), as shown in Equation [2].

$$\mathbb{E}(\widehat{f_2}) = \mathbb{E}\left(100 - 25 \cdot log_{10}\left(1 + (1/p)\sum_{i=1}^{p}(\bar{X}_{Ti} - \bar{X}_{Ri})^2\right)\right)$$

$$\approx 100 - 25 \cdot log_{10}\left(1 + \mathbb{E}\left((1/p)\sum_{i=1}^{p}(\bar{X}_{Ti} - \bar{X}_{Ri})^2\right)\right)$$

$$= 100 - 25 \cdot log_{10}\left(1 + (1/p)\left(\sum_{i=1}^{p}(\mu_{Ti} - \mu_{Ri})^2 + \sum_{i=1}^{p}(\sigma_{Ti}^2 + \sigma_{Ri}^2)/n\right)\right)$$

**Equation [2]: Approximation of the expected value of $\widehat{f_2}$ from Shah et al. (1998). Note the 'approximately equal' sign in the second line.**

The expected value of $\widehat{f_2}$ has the form

$$100 - 25 \cdot log_{10}(1 + g(\boldsymbol{\mu}) + \frac{h(\sigma^2)}{n}),$$

where $\boldsymbol{\mu} = (\boldsymbol{\mu}_R, \boldsymbol{\mu}_T)$ is the vector of expected values and $\boldsymbol{\sigma}^2 = (\boldsymbol{\sigma}_R^2, \boldsymbol{\sigma}_T^2)$ is the vector of variances. The auxiliary functions *g()* and *h()* are introduced to simplify the formulae and to show the relationships between the suggestions of Shah et al. (1998) and the suggestions of the authors on $\hat{f}_{2,\text{exp}}$ (see next page).

Under the approximation shown in Equation [2], the bias depends on $h(\boldsymbol{\sigma}^2)/n$ and the point estimate $\widehat{f_2}$ is asymptotically unbiased.

Shah et al. (1998) suggest **an approximation** to reduce the bias by subtracting the estimate $h(\hat{\boldsymbol{\sigma}}^2)/n$ from the argument in the logarithm in the formula for $\widehat{f_2}$:

$$\hat{f}_{2,bc} = 100 - 25 \cdot log_{10}\left(1 + g(\hat{\boldsymbol{\mu}}) - \frac{h(\hat{\boldsymbol{\sigma}}^2)}{n}\right)$$

$$= 100 - 25 \cdot log_{10}(1 + (1/p)\left[\sum_{i=1}^{p}(\bar{X}_{Ti} - \bar{X}_{Ri})^2 - 1/n\sum_{i=1}^{p}(S_{Ti}^2 + S_{Ri}^2)\right]).$$

Shah et al. (1998) do not claim that this is an unbiased estimate; they describe it as an "**intuitive bias correction**". This bias correction is **not always applicable** because the argument of the logarithm may be negative. This can happen when the squared ED between profile means is smaller than the total variance divided by the sample size, e.g. when the data are highly variable or the mean differences are small.

Recall the role of the auxiliary functions *g()* and *h()* introduced above with
$g(\hat{\boldsymbol{\mu}}) = (1/p)\sum_{i=1}^{p}(\bar{X}_{Ti} + \bar{X}_{Ri})^2$ and $h(\hat{\boldsymbol{\sigma}}^2) = (1/p)\sum_{i=1}^{p}(S_{Ti}^2 + S_{Ri}^2)$.
We now want to show the relationship between the "intuitive bias correction" $\hat{f}_{2,bc}$ suggested by Shah et al. (1998) and the "expected $f_2$" as suggested in Noce et al. (2020) and Xu et al. (2021).

"Intuitive bias correction" $\quad \hat{f}_{2,bc} = 100 - 25 \cdot log\left(1 + g(\hat{\boldsymbol{\mu}}) - \frac{h(\hat{\boldsymbol{\sigma}}^2)}{n}\right)$

"Expected $f_2$" $\quad \hat{f}_{2,exp} = 100 - 25 \cdot log\left(1 + g(\hat{\boldsymbol{\mu}}) + \frac{h(\hat{\boldsymbol{\sigma}}^2)}{n}\right)$

**Equation [3]: Relationship between "intuitive bias correction" $\hat{\boldsymbol{f}}_{\boldsymbol{2,bc}}$ and "Expected $\boldsymbol{f_2}$".**

For the so-called "expected $f_2$" ($\hat{f}_{2,\mathrm{exp}}$), it is suggested that $h(\hat{\boldsymbol{\sigma}}^2)/n$ should be added to the argument of the logarithm instead of subtracting it (see red arrow). No justification is provided for this approach, nor is a reference given where this suggestion can be found.
We distinguish between the expected value $\mathbb{E}(\hat{f}_2)$ as derived by Shah et al. (1998), a theoretical property of the estimator $\hat{f}_2$ expressed in terms of the population quantities $\boldsymbol{\mu} = (\boldsymbol{\mu_R}, \boldsymbol{\mu_T})$ and $\boldsymbol{\sigma}^2 = (\boldsymbol{\sigma}_R^2, \boldsymbol{\sigma}_T^2)$ , and the sample-level quantity $\hat{f}_{2,\mathrm{exp}}$ as computed by Noce et al. (2020). $\hat{f}_{2,\mathrm{exp}}$ is obtained by replacing $\boldsymbol{\mu}$ and $\boldsymbol{\sigma}^2$ with their sample counterparts in the formula for the approximated expected value $\mathbb{E}(\hat{f}_2)$ (Equation [4]).

$$E(f_2) = 50 \cdot log\left\{\left[1 + \frac{1}{P}\sum_{j=1}^{P}(R_j - T_j)^2 + \sum_{j=1}^{P}(s_{Rj}^2 + s_{Tj}^2)/n\ )\right]^{-0.5} \cdot 100\right\}$$

**Equation [4]: Formula as published in Noce et al. (2020, Eq. (3)), retaining the notation used in that work. The orphaned closing bracket (highlighted in orange by present authors) has no corresponding opening bracket.**

Please note that we prefer the notation $\hat{f}_{2,\mathrm{exp}}$ for the methods presented in Equations [4] and [5] because the notations 'E($f_2$)' or 'expected $f_2$' as used by Noce et al. (2020) and Xu et al. (2021) are misleading. The latter notations falsely suggest that a computable sample statistic is an expected value, when in fact $\mathbb{E}(\hat{f}_2)$ as presented in Equation [2] is a theoretical property of the sampling distribution that cannot be calculated from limited data.

Noce et al. (2020) introduce this sample-level computation citing Shah et al. (1998) and Ma et al. (1999), but neither of these references provides a justification for replacing the population quantities with sample analogs. As explained above, Shah et al. (1998) and Ma et al. (1999) derive $\mathbb{E}(\hat{f}_2)$ as a theoretical quantity and use the term 'expected value of $\hat{f}_2$' exclusively in the statistical sense of an expected value of the sampling distribution, not in the sample-level interpretation adopted by Noce et al. (2020).

Xu et al. (2021) adopt a similar sample-level computation, citing Ma et al. (2000) and Mendyk et al. (2013). None of these references provides a justification for this substitution. A third mentioned reference is incomplete.

## 6 Formula ambiguity in Noce et al. (2020)

Beyond the concerns outlined above, the formula for $\hat{f}_{2,\text{exp}}$ as published by Noce et al. (2020) contains a notation ambiguity. Equation [4] (Noce et al. (2020)) and Equation [5] (Xu et al. (2021)) show the formulas in the respective manuscripts.

$$\hat{f}_{2,exp} = 100 - 25 \cdot log_{10}\left(1 + \frac{1}{p}\left(\sum_{i=1}^{p}\left(\bar{X}_{T,i} - \bar{X}_{R,i}\right)^2 + \frac{1}{n}\sum_{i=1}^{p}\left(S_{T,i}^2 + S_{R,i}^2\right)\right)\right)$$

**Equation [5]: Suggestion of “expected f2” as proposed by Xu et al. (2021).**

In Equation [5] (Xu et al. (2021)), the bracket structure is unambiguous: “$1/p$” refers to both the squared Euclidean distance between the profile means and to the sum of estimated variances divided by the sample size. In Equation [4] (Noce et al. (2020)), the orphaned closing bracket leaves the intended grouping unclear. Depending on where the missing opening bracket is placed, the formula either coincides with Equation [5] or defines a different quantity with different numerical results and – at times – different conclusions. As it stands, the ‘*1/p*’ only refers to the first sum. As no mathematical derivation accompanies the formula, the intended reading cannot be determined. This ambiguity has propagated into the regulatory guidance: the version published in Noce et al. (2020) is implemented by the South

Korean MFDS guideline (2023), whereas the variant published by Xu et al. (2021) was implemented in the EMA Q&A 3.13 (2025). Note that the EMA references Noce et al. (2020), but the formula quoted originates from Xu et al. (2021).

## 7 Statistical consequences of the variance term reversal

Recall the two suggested point estimates from Shah et al. (1998) and the $\hat{f}_{2,\mathrm{exp}}$ authors (Noce et al., (2020); Xu et al., (2021)) in Equation [3]. The effect of adding the term $h(\hat{\boldsymbol{\sigma}})/n$ to the argument of the logarithm is that the $\hat{f}_{2,\mathrm{exp}}$ method becomes more conservative compared to other approaches:

1) The T1E rate is, of course, lower than the T1E rate of other approaches. This property is claimed by Noce et al. (2020) and Xu et al. (2021) as an advantage.
2) For highly variable profiles, the arbitrary term $h(\hat{\boldsymbol{\sigma}})/n$ becomes **dominant** in the decision-making process. As a consequence, the power of the $\hat{f}_{2,\mathrm{exp}}$ method will not increase above a certain limit for highly variable profiles even if the true differences between the profile means tend to zero. This behaviour is not a property of a well-constructed test statistic, and it has no mathematical justification.

Please also note that $\hat{f}_{2,\mathrm{exp}}$ can only be used for equal samples sizes of REF and TEST products which is in practice often but not always fulfilled. This also implies that the bias-corrected and accelerated (BCa) bootstrap interval cannot be calculated because the jackknife step within the bootstrap simulation leads to unequal sizes of the bootstrap samples.

## 8 Numerical illustration: behavior of $\hat{f}_{2,\mathrm{exp}}$ under equivalence

Figure [1] presents an example evaluation of a highly variable dissolution profile dataset. Recall the equivalence hypotheses presented in Equation [1]. They describe that equivalence should be claimed if the quadratic mean of the profile mean differences (QMD) is below 10% (of Label Claim). In the example evaluation the REF population is compared to itself: REF and TEST samples are identical, the difference is 0 at all dissolution time points resulting in $QMD = 0$. An equivalence procedure may be expected to decide in favour of equivalence in such situations. The point estimate $\widehat{f_2}$ results in the maximum value 100. The bootstrapped lower 95% confidence limits result in the values 35.2 (when disregarding the orphaned

bracket in the ambiguous Noce et al. (2020) notation) and 43.3 (using the unambiguous formula from Xu et al. (2021)). Both $\hat{f}_{2,\text{exp}}$ variants decide against equivalence. As explained in section 7 “Statistical consequences of the variance term reversal” the large estimated variances dominate the equivalence decision via the term $h(\hat{\boldsymbol{\sigma}})/n$. This leads to an extreme difference between the point estimate and the respective lower confidence limit. Obviously, the maximum power obtained for both $\hat{f}_{2,\text{exp}}$ variants is close to 0 for products with such dissolution profile variability.

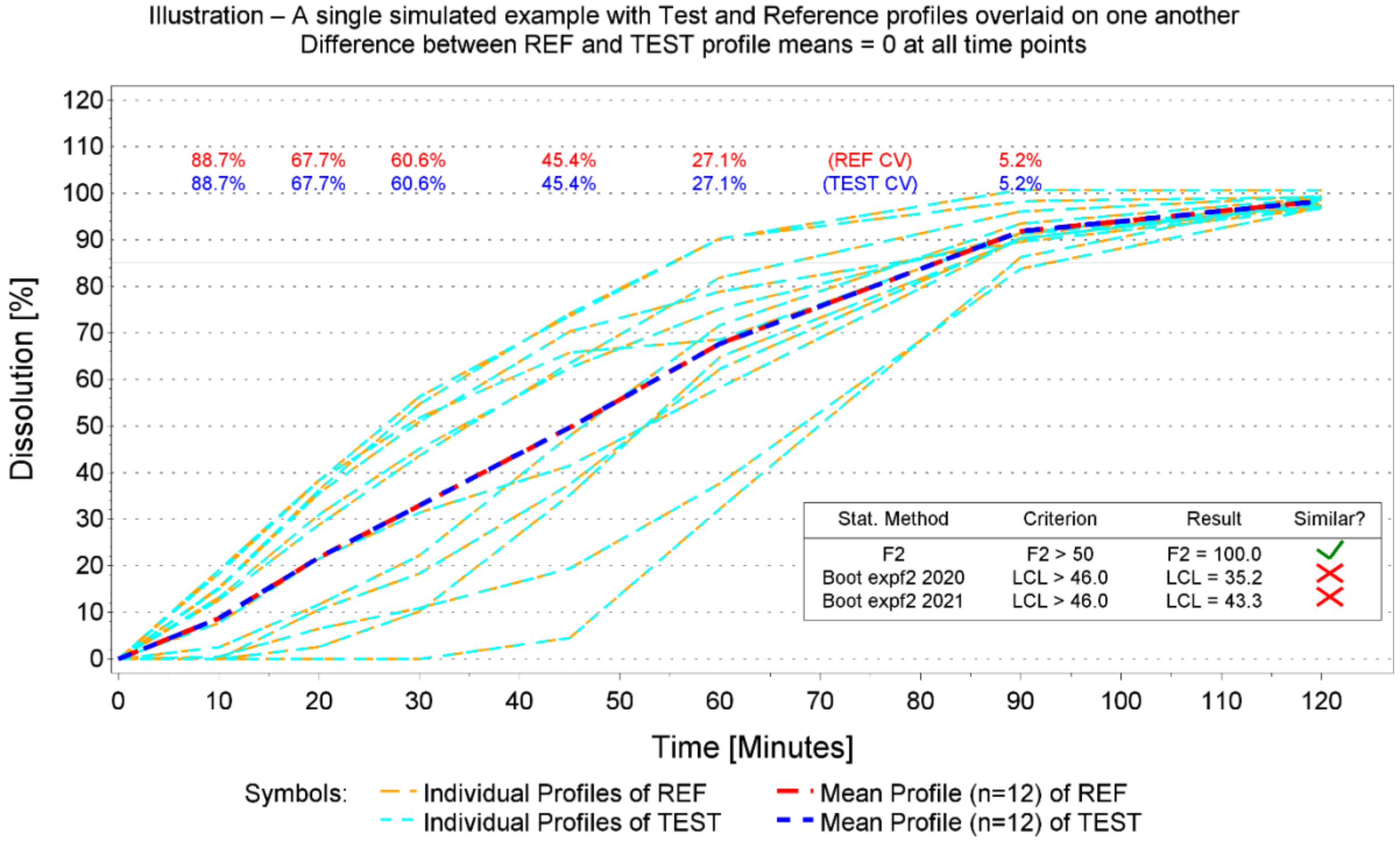

**Figure [1]: Example evaluation of a simulated, highly variable dissolution profile dataset. Reference and test dataset coincide.**

## 9 Assessment by FDA statisticians

As the authors from Shah et al. (1998) are from FDA, the claim that $\hat{f}_{2,\text{exp}}$ is suggested by them brings a lot of attention for such statements. More recently, FDA statisticians Liu et al. (2024) clarified that $\hat{f}_{2,\text{exp}}$ is not mathematically justifiable. Regarding the statements in Xu et al. (2021), Liu et al. (2024) clarify:

> “*Their paper includes three additional $f_2$ estimators ($\hat{f}_{2,vcbc}$, $\hat{f}_{2,exp}$ and $\hat{f}_{2,vcexp}$) for correcting the bias or variance of the naive $f_2$ estimators. However, we do not recommend those methods because their proposed variance correction for the $f_2$* *estimators is not mathematically justifiable, and their proposed expected $f_2$ estimators correct the bias to the opposite direction (i.e. further increase the bias of the naive $f_2$ estimators).”*

## 10 Working group survey on perceived advantages and disadvantages

A survey was conducted among working group members to systematically capture perceived advantages and disadvantages of $\hat{f}_{2,\text{exp}}$. Respondents consistently struggled to identify advantages. The most frequently cited property was the method's conservatism, which was viewed as double-edged: while it may ease regulatory acceptance when similarity is demonstrated, it was equally considered a substantial limitation due to the resulting loss of statistical power. Several respondents stated explicitly that they see no advantages whatsoever. On the disadvantage side, concerns were both fundamental and practical: the lack of a sound mathematical derivation, low statistical power with high-variability data, the notation ambiguity in the formula as published by Noce et al. (2020), and the fact that the method can reject equivalence even when mean differences are zero at all time points. The growing adoption by health authorities beyond EMA was noted as a particular concern, given the risk of entrenching a fundamentally flawed method in regulatory practice.

## 11 Summary

This paper identifies four fundamental concerns with the so-called 'expected $f_2$' ($\hat{f}_{2,\text{exp}}$).

First, the formula has no traceable origin in the references cited by its proponents. Noce et al. (2020) and Xu et al. (2021) attribute $\hat{f}_{2,\text{exp}}$ to Shah et al. (1998) and Ma et al. (1999, 2000), but a review of these sources shows that neither mentions nor suggests the sample-level computation for comparing dissolution profiles.

Second, no mathematical justification has been provided for the $\hat{f}_{2,\text{exp}}$ formula. This has been confirmed by FDA statisticians Liu et al. (2024), who stated that the variance correction is “not mathematically justifiable” and that the expected $f_2$ estimators “correct the bias to the opposite direction (i.e. further increase the bias of the naive estimators)”.

Third, the method exhibits poor statistical properties: for highly variable profiles, the variance term dominates the statistic, resulting in power that does not increase even as the true difference between profiles approaches zero. The method can reject equivalence when profiles are identical.

Fourth, the formula as published by Noce et al. (2020) contains a notation ambiguity that renders the intended grouping of terms unclear. This ambiguity has propagated into regulatory guidance.

A survey of working group members, designed to elicit arguments both for and against the method, found no scientifically meaningful advantage.

## 12 Conclusion

At a follow-up meeting on 24 February 2026, the working group jointly concluded that $\hat{f}_{2,\text{exp}}$ should not be recommended for dissolution profile comparison. The group agreed to document its assessment in this position paper and to engage with respective health authorities to advocate against the continued use of this method.

We note that the broader methodological landscape for dissolution profile comparison is well developed. The 2019 M-CERSI workshop on dissolution profile similarity, attended by approximately 150 participants from the CMC area, produced a series of publications including decision trees for selecting appropriate equivalence tests depending on product characteristics. Scientifically sound alternatives to $\hat{f}_{2,\text{exp}}$ exist and are published. Reference is made to Suarez-Sharp et al. (2020) and Hoffelder et al. (2022).

ICH M13B (2025) offered an opportunity to establish a harmonized, statistically grounded framework for dissolution profile comparison. CMC statisticians were consulted during its development, producing recommendations that aligned with the M-CERSI workshop outcomes but were not reflected in the final draft. The EFSPI CMCSNE SIG working group hopes that future revisions will be more closely aligned to the M-CERSI workshop outcomes.

The EFSPI CMCSNE SIG, comprising of statisticians and allies from more than twenty five European based organisations, stands ready to contribute to this process and welcomes dialogue with health authorities, ICH working groups and the wider CMC community.

We are open for debate.

## 14 Version

This version:

- 2026-06-01 Consistency checks of equations and nomenclature. Added references for M-CERSI workshop.

Previous versions:

- 2026-05-20 Added "Participants of Working Group" section with Contributors and Observer. Added disclaimer on contributions by public servants. Updated Figure [1] title. Reframed Section 6: "formula inconsistency" replaced by "formula ambiguity." Equation [4] replaced with annotated screenshot from Noce et al. (2020). Corresponding adjustments in Abstract, Summary, Survey and Example sections. Added disclosure on shareholdings.
- 2026-04-29 Version for IP Clearance by contributors.